\documentclass{article}

\PassOptionsToPackage{numbers, sort&compress}{natbib}
\usepackage[final]{neurips_2021_ml4ps}

\usepackage[pdftex]{graphicx} 
\usepackage[utf8]{inputenc} 
\usepackage[T1]{fontenc}    
\usepackage[colorlinks, citecolor=green!80!black, urlcolor=blue, linkcolor=red]{hyperref}
\usepackage{url}            
\usepackage{booktabs}       
\usepackage{amsfonts,amsmath,amssymb}       
\usepackage{nicefrac}       
\usepackage{microtype}      
\usepackage{xcolor}         

\title{Neural quantum states \\for supersymmetric quantum gauge theories}

\author{%
   Xizhi Han \\
   Kavli Institute for Theoretical Physics \\
   Santa Barbara, CA, US \\
   \texttt{hanxzh@ucsb.edu} \\

   \And
  Enrico Rinaldi \\
  University of Michigan, Ann Arbor, MI, US \\
  and Theoretical Quantum Physics Laboratory, RIKEN, Japan \\
  and iTHEMS, RIKEN, Japan \\
  \texttt{erinaldi@umich.edu} \\
}

\begin{document}

\maketitle

\begin{abstract}
    Supersymmetric quantum gauge theories are important mathematical tools in high energy physics.
    As an example, supersymmetric matrix models can be used as a holographic description of quantum black holes.
    The wave function of such supersymmetric gauge theories is not known and it is challenging to obtain with traditional techniques.
    We employ a neural quantum state ansatz for the wave function of a supersymmetric matrix model and use a variational quantum Monte Carlo approach to discover the ground state of the system.
    We discuss the difficulty of including bosonic particles and fermionic particles, as well as gauge degrees of freedom.
\end{abstract}

\section{Introduction}\label{sec:introduction}

The powerful conjecture of gauge/gravity duality~\cite{Maldacena:1997re,Itzhaki:1998dd} translates difficult (or intractable) problems in quantum gravity to well-defined problems in non-gravitational quantum theories.
Quantum mechanics of matrices (or matrix models in the following) can provide us with a nonperturbative formulation of superstring/M-theory~\cite{Banks:1996vh,deWit:1988wri,Itzhaki:1998dd,Berenstein:2002jq}: at strong coupling, and in the limit of large $N\times N$ matrices, weakly-coupled gravity with small stringy corrections can be described by matrix models of Yang--Mills-type (with gauge degrees of freedom).
A bound state of D0-branes with open strings can be interpreted as a black hole and described by a supersymmetric quantum field theory with bosonic, fermionic, and gauge degrees of freedom.
On the other hand, such bound state of $N$ D0-branes has a dynamics accurately described by quantum mechanics of large $N \times N$ traceless Hermitean matrices.

The ground state wave function of complicated matrix models, such as the aforementioned ones arising from dimensional reduction of supersymmetric SU($N$) Yang--Mills theories, is not known.
However, inspecting the wave function's structure is of paramount importance to understand quantum information concepts and one interesting application is the emergence of space-time geometry from the entanglement of quantum information in a holographic approach~\cite{Hanada:2021ipb}.
Unfortunately, traditional methods used to solve matrix models, such as Lattice Monte Carlo simulations of the Euclidean path integral~\cite{Berkowitz:2016jlq}, can not access the wave function, and are challenging for systems with a sign problem, where the Boltzmann weight for each path can not be interpreted as a positive probability measure~\cite{Alford:1998sd}.
Other numerical approaches to quantum states include quantum computing and tensor network methods.
The former is currently limited by the small number of physical qubits and the latter are unlikely to yield an advantage in this case because matrix models lack the spatial geometric locality which is present in lattice spin models.

Deep artificial neural network architectures, routinely used in the context of density estimation, can be used to efficiently approximate quantum states, without imposing constraining structures on the wave function or limiting their representation power by focusing on simple distributions.
Recently, different numerical approaches to matrix model wave functions have been compared in \cite{rinaldi2021matrix}.

\section{Related works}\label{sec:related}

There has been tremendous work on studying neural network wave function ansatz for quantum many-body systems~\cite{glasser2018neural, carleo2019machine}.
Compared to spin models, neural quantum states of bosons and fermions are less well understood, and previous research mostly focused on representing the state in the particle number (Fock) basis~\cite{saito2018machine, li2019accelerating, nomura2020machine}.
In the Fock basis, the number of bosons or fermions in each mode is a discrete variable, and hence can be processed in similar ways as a quantum spin.
An interesting proposal of continuous variable RBM states was discussed in Ref.~\cite{stokes2021continuous}.
In our work, we employ generative flows as variational wave functions without truncating the bosonic modes.
Bosons are represented in their physical infinite-dimensional Hilbert space.
We observe both lower variational energies and better symmetry properties compared to the Fock basis methods.

The generative variational wave functions in this work are accompanied by generative flow Monte Carlo samplers for improved efficiency.
The energy of the variational wave function is often evaluated with Monte Carlo methods, for example with a Gibbs sampler.
Recently, neural generative models have shown advantages over the conventional methods, both in spin \cite{sharir2020deep, vielhaben2021generative, mcnaughton2020boosting, wu2019solving} and field systems \cite{albergo2019flow, medvidovic2020generative, nicoli2021estimation, kanwar2020equivariant}.
We use the Block Neural Autoregressive Flow (BNAF)~\cite{decao2019block} as the Monte Carlo sampler, which generates uncorrelated bosonic and fermionic configurations.

Preservation of gauge symmetries is essential for physical interpretations of the resulting variational state.
Recent success in designing symmetry-invariant or equivariant neural states~\cite{Han:2019wue, luo2020gauge, luo2021gauge, favoni2020lattice, kanwar2020equivariant, kohler2020equivariant} has shed light on this difficult task.
Research for equivariant variational states with non-Abelian symmetries, or even supersymmetries, however, is still incomplete.
In this work the gauge symmetry is not imposed exactly.
Instead, we present two general penalty and projection methods for finding gauge-invariant states, and observe satisfactory results.
In the supersymmetric case supersymmetries (or more generally, the BPS conditions~\cite{Witten:1978mh}) are also preserved to high accuracy.

\section{Methods}\label{sec:methods}

\subsection{Variational Quantum Monte Carlo}\label{sec:variational-montecarlo}

The variational quantum Monte Carlo method consists of three components: a wave function network, a sampler network, and an optimizer.
The wave function network is a complex function $\psi_\theta(X)$ parametrized by $\theta$.
Here $X$ is the input vector denoting a basis state, for example a binary vector of spins ups and downs, or a real vector of coordinates of the particles.
The sampler is a generative model capable of evaluating and sampling from a probability distribution $p_{\eta}(X)$ ($\eta$ is the vector of parameters), where the samples can be fed into the wave function network.
The optimizer updates the parameters $\theta$ and $\eta$ in the networks to lower the variational energy.

The variational energy is computed from Monte Carlo samples: for a Hamiltonian operator $\hat{H}$, the variational energy $E_\theta$ is
\begin{align}\label{eq:variational-mc}
E_\theta \equiv
\langle\psi_\theta|\hat{H}|\psi_\theta\rangle
=
\int dX |\psi_\theta(X)|^2\cdot\frac{\langle X |\hat{H} |\psi_\theta\rangle}{\psi_\theta(X)} = \mathbf{E}_{X \sim p_\eta}\left[\frac{\epsilon_\theta(X) |\psi_\theta(X)|^2}{p_\eta(X)}\right]
\ .
\end{align}
Here $\epsilon_\theta(X)$ is defined as $\langle X | \hat{H} | \psi_\theta \rangle / \psi_\theta(X)$, and the expectation value is estimated as the mean of the samples.

During training, the parameters $\theta$ are updated to minimize the energy $E_\theta$ and the sampler parameters $\eta$ are updated to minimize the  Kullback–Leibler divergence between distributions $p_\eta$ and $|\psi_\theta|^2$:
\begin{equation} \label{eq:dkl}
    D_{\textrm{KL}}(p_\eta \| |\psi_\theta|^2) = \mathbf{E}_{X \sim p_\eta}\left[\log \frac{p_{\eta}(X)}{|\psi_\theta(X)|^2}\right].
\end{equation}
In particular, during each step we assign $\theta \leftarrow \theta - \beta \nabla_\theta E_\theta$, and $\eta \leftarrow \eta - \beta \nabla_\eta D_{\textrm{KL}}(p_\eta \| |\psi_{\theta}|^2)$, with a learning rate $\beta = 10^{-3}$. The gradients are
\begin{equation}
    \nabla_\theta E_\theta = \mathbf{E}_{X \sim p_\eta}\left[\frac{\nabla_\theta (\epsilon_\theta(X) |\psi_\theta(X)|^2)}{p_\eta(X)}\right],
\end{equation}
\begin{equation}
    \nabla_\eta D_{\textrm{KL}}(p_\eta \| |\psi_\theta|^2) = \mathbf{E}_{X \sim p_\eta}\left[\nabla_\eta( \log p_\eta) \log \frac{ p_{\eta}(X)}{|\psi_\theta(X)|^2}\right], \label{eq:grad_DKL}
\end{equation}
and the expectation values $\mathbf{E}[f(X)]$ are estimated as Monte Carlo sample averages $\sum_{i = 1}^K f(X_i) / K$ in stochastic gradient descents. In (\ref{eq:grad_DKL}), $\nabla_\eta (\log p_\eta)$ comes from a gradient of the sampling distribution $X \sim p_\eta$, and we have also noted that $\mathbf{E}_{X \sim p_\eta}[\nabla_\eta \log p_\eta(X)] = 0$. The (locally) minimal value for $E_\theta$ found is a variational upper bound for the ground state energy.

We have chosen to minimize $D_{\textrm{KL}}(p_\eta \| |\psi_\theta|^2)$ to reduce the variance of the Monte Carlo estimate for (\ref{eq:variational-mc}).
This choice is guided by the following consideration. In the limit where both the wavefunction and the sampler networks have strong representation power, $p_\eta = |\psi_\theta|^2$ minimizes the KL divergence (\ref{eq:dkl}) and $\psi_\theta = \psi_0$ minimizes the energy $E_\theta$, where $\hat{H} |\psi_0\rangle = E_0 |\psi_0\rangle$ is the true ground state wavefunction.
In this limit, the quantity in the expectation value of (\ref{eq:variational-mc}) is a constant $E_0$, and hence its Monte Carlo estimate has zero variance.
Hence in practice we expect and see small and controlled sample variance for the estimate of (\ref{eq:variational-mc}), showing the effectiveness of this procedure.

\subsection{Wave function and sampler}\label{sec:neural-ansatz}

Generative flows are essential in both our wave function and sampler networks.
The basic idea of flow models is to generate complex distributions by applying nonlinear but reversible transformations to simpler ones.
If $x$ is a sample from the (continuous) distribution $p_X(x)$, and $y = F(x)$, by a change of variable we know that $p_Y(y) = p_X(F^{-1}(y)) |\mathrm{det} D F|^{-1}$.
The resulting $p_Y(y)$ can be very complicated if $F$ is highly nonlinear.
We can then use feed-forward networks as the transformation $F$, but we should guarantee that $F$ is reversible and the Jacobian is efficient to compute.

In the BNAFs, reversibility is guaranteed by imposing some autoregressive structure.
Basically if $y$ and $x$ are vectors, $y_i$ depends only on $x_1$ to $x_i$, and $\partial y_i / \partial x_i > 0$, the Jacobian would be lower triangular with positive diagonals, and hence invertible.
Its determinant $\prod_i \partial y_i / \partial x_i$ is also easy to compute.

Let $p_\theta(X)$ be a probability distribution constructed from the BNAF.
In the bosonic case, our variational wave function ansatz is then $\psi_{\theta}(X) = \sqrt{p_{\theta}(X)} \exp(i F_\theta(X))$, where $F_\theta$ is a fully-connected feed-forward network generating the phase.
The sampler could just be $p_\theta$, sharing the parameters, or a separate BNAF $p_\eta$.

For supersymmetric models, the quantum wave function $\psi(X, \xi)$ depends on bosonic matrices $X$ and fermionic matrices $\xi$.
For $\xi$ we choose the fermion number basis as the basis for quantum states: the coordinates $(X, \xi)$ for the wave function contain real numbers for $X$ and binaries for $\xi$.
In this case we use $\sqrt{p_{\theta}(X)} (f_\xi(X) + i g_\xi(X))$ as the variational wave function, where $p_{\theta}(X)$ is given by a BNAF and $f_\xi$ and $g_\xi$ are real functions given by fully-connected neural networks.
The weights and biases in $f_\xi$ and $g_\xi$ depend on the binary vector $\xi$, via additional fully-connected networks. Because of the use of $f_\xi$ and $g_\xi$, the sampler is a separate BNAF, and in general will be different from $p_\theta$.

In more detail, we implement the same BNAF architecture as described in Ref.~\cite{decao2019block}, with default hyperparameter choices as follows.
For each BNAF network we have one hidden layer. The activation function is \texttt{SinhArcsinh} with trainable skewness and tailweight parameters.
The number of hidden units is $\alpha$ times the number of input (or output) units.
Results for different $\alpha$'s are compared in Tables \ref{tab:ml_bosonic_su2} and \ref{tab:ml_susy_su2}.
We have also experimented with other generative flow architectures~\cite{Han:2019wue}, and BNAF shows the best accuracy in the models we consider.

\subsection{Preservation of gauge symmetries}

In this work gauge invariance is imposed by either adding a gauge Casimir penalty, or by projecting the state onto the gauge invariant subspace, during or after training.
A gauge Casimir penalty can be added as a perturbation to the Hamiltonian,
$
\hat{H}' = \hat{H} + c\sum_\alpha\hat{G}_\alpha^2
$, where $c > 0$ is a hyperparameter and $\hat{G}_\alpha$ are operators generating the gauge symmetry.
The projection $\hat{P}$ onto the singlet sector can be written as an average over gauge transformations
$
\hat{P} |\psi\rangle = \int d U \, \hat{U} |\psi\rangle
$, where the integration is over the gauge group.
For any gauge invariant observable $\hat{O}$, its expectation value in the projected state is
$
\langle \hat{O} \rangle_{\text{singlet}} = \langle \psi | \hat{P} \hat{O} | \psi \rangle / \langle \psi | \hat{P} | \psi \rangle
$, and can be estimated with Monte Carlo samples of $U$ and $X$.

\section{Results}\label{sec:results}

We consider two matrix models to benchmark our method: a boson-only matrix model with 2 matrices and SU(2) gauge group (hereafter, the bosonic model), and a supersymmetric matrix model with 2 bosonic matrices and 1 fermionic matrix with SU(2) gauge group (hereafter, the minimal BMN model~\cite{Berenstein:2002jq}).
Both models can be defined with local Hamiltonians described in details by Ref.~\cite{rinaldi2021matrix} and the sought-after physical states are those invariant under SU(2) gauge transformations.

Our benchmark focuses on the expectation value of relevant operators (energy, gauge Casimir) for various strengths of the gauge coupling $\lambda = g^2 N = 2 g^2$.
For these SU(2) gauge models, where the number of degrees of freedom is limited, we can compare directly with results from exact diagonalization of a truncated Hamiltonian (using the Fock basis) in the limit where the truncation effects are negligible~\cite{rinaldi2021matrix}.
Moreover, for the supersymmetric model we know that the energy of the ground state is exactly zero at each coupling due to supersymmetry.

All results in this section are obtained by training the parameters $(\theta, \eta)$ on a multi-core CPU laptop in less than 24 hours, with 200 samples in each mini-batch, 1000 batches in each epoch, and about 150 training epochs for each coupling.
In training, the Adam optimizer is used with an initial learning rate $\beta = 10^{-3}$ and the rate is reduced when the loss function plateaus.
The training is stopped when the learning rate drops below $10^{-5}$.
The final expectation value of observables is evaluated from $10^6$ Monte Carlo samples.

\begin{table}[ht]
    \caption{Neural variational quantum Monte Carlo ground state energy for the bosonic model.
    The exact result of the Hamiltonian truncation (HT) is reported in the last column.
    }
    \label{tab:ml_bosonic_su2}
    \centering
    \begin{tabular}{c|ccccccc}
    \toprule
     $\alpha$ & 1 & 2 & 5 & 10 & 20 & 50 & HT (exact) \\
    \midrule
    $\lambda =$ 0.2 & 3.137(2) & 3.137(2) & 3.140(2) & 3.138(2) & 3.137(2) & 3.135(2) & 3.134 \\
    $\lambda =$ 0.5 & 3.313(2) & 3.312(2) & 3.308(2) & 3.307(2) & 3.302(2) & 3.305(2) & 3.297\\
    $\lambda =$ 1.0 & 3.544(3) & 3.544(2) & 3.541(3) & 3.528(2) & 3.519(2) & 3.520(2) & 3.516\\
    $\lambda =$ 2.0 & 3.914(3) & 3.910(3) & 3.892(3) & 3.872(3) & 3.857(3) & 3.859(3) & 3.854\\
    \bottomrule
    \end{tabular}
\end{table}

For the SU(2) bosonic model, as a warm up test for our neural ansatz, we assess the accuracy of the variational energies by increasing the ratio $\alpha$ of the number of hidden units to features in the BNAF $\psi_\theta(X)$.
With more parameters (larger $\alpha$) we expect that the ansatz is more flexible and accurate, and indeed convergence to exact results for the energy of the ground state $E_0$ is observed in Table~\ref{tab:ml_bosonic_su2} for $\alpha \ge 20$.
More parameters are needed to represent the ground state accurately at larger gauge coupling strengths (for reference, the zero coupling limit of this model is just a system with 6 independent harmonic oscillators).
We have also tested our neural ansatz on the case of larger SU($N$) algebras, which amounts to having a larger number of continuous variables as input to the neural network, without any appreciable degradation of the accuracy of the results, or an inflation of the training time.

In the supersymmetric model we see vanishing variational energies as we increase $\alpha$, as expected.
In Table~\ref{tab:ml_susy_su2} we also report the convergence of other observables: the gauge Casimir $G^2 = \sum_\alpha G_\alpha^2$, the SO(2) angular momentum $M$, and the fermion number $F$.
When evaluated on the variational state, they also converge to correct values for $\alpha \ge 20$.
For example, the ground-state energy of bosonic and fermionic oscillators are $\pm(N^2-1)=\pm 3$, and they cancel out.
The results at $\alpha = 20$ means the cancellation is reproduced up to one-percent-order error.

\begin{table}[ht]
    \caption{
    Neural variational quantum Monte Carlo ground state observables for the SU(2) minimal BMN model, at coupling $\lambda = g^2 N = 1.0$ and $\mu = 1$.
    The exact result of the Hamiltonian truncation (HT) is reported in the last column.
    }
    \label{tab:ml_susy_su2}
    \centering
    \begin{tabular}{c|cccccc}
    \toprule
     $\alpha$ & 1 & 5 & 10 & 20 & 50 & HT (exact)\\
    \midrule
    $H$   & 0.058(6) &  0.041(6) & 0.031(6) & 0.014(6) & 0.005(6) & 0.000 \\
    $G^2$ & 0.007(8) &  0.014(8) & 0.007(9) & 0.022(9) & 0.012(9) & 0.000 \\
    $M$   & -0.0003(3) &  -0.0001(4) & 0.0001(4) & -0.0003(5) & -0.0001(4) & 0.0000 \\
    $F$   & 0.1844(6) & 0.1895(6) & 0.1922(6) & 0.1946(7) & 0.1935(7) & 0.2034 \\
    \bottomrule
    \end{tabular}
\end{table}

Results at $\alpha = 20$ for other couplings are summarized in Table~\ref{tab:ml_susy_su2_couplings} and we can see that they become less precise as the system becomes more strongly coupled (note change in $F$ for different $\lambda$ is expected).

\begin{table}[ht]
    \caption{
    Neural variational quantum Monte Carlo ground state observables for the SU(2) minimal BMN model at various couplings $\lambda$.
    }
    \label{tab:ml_susy_su2_couplings}
    \centering
    \begin{tabular}{c|ccc}
    \toprule
    $\lambda$ & 0.5 & 1.0 & 2.0 \\
    \midrule
    $H$ & 0.009(5) & 0.014(6) & 0.034(7) \\
    $G^2$ & 0.010(6) & 0.022(9) & 0.038(14) \\
    $M$ & -0.0002(3) & -0.0003(5) & 0.0006(7) \\
    $F$ & 0.1224(4) & 0.1946(7) & 0.2729(9) \\
    \bottomrule
    \end{tabular}
\end{table}

\section{Conclusions}\label{sec:conclusions}

In this work we have shown how effective a BNAF variational quantum state is in a supersymmetric model with bosonic, fermionic, and gauge degrees of freedom.
The ground state properties are correctly reproduced when enough parameters are included in the neural ansatz using continuous variables to represent bosons.
We note how this ansatz is more limited when the system is driven to stronger coupling where the number of fermions and bosons can increase and the wave function becomes more complicated.

As a remark, we think that these supersymmetric matrix models could become challenging benchmark tasks for developing new neural ansatz architectures and algorithms.
On such benchmark tasks one could tune the difficulty by choosing the number of bosons (and even add fermions) and the gauge group algebra, and some tasks also have exact results or analytical predictions (e.g. at very large $N$). At large $N$ and strong coupling, the wave function should contain details of quantum gravity, and hence an efficient numerical method would be both powerful and exciting.

\begin{ack}
We thank Masanori Hanada, Michael McGuigan, Franco Nori, Mohammad Hassan, and Yuan Feng for collaboration on a related work.
E.~R. is supported by Nippon Telegraph and Telephone Corporation (NTT) Research.
We gratefully acknowledge the use of the Supercomputer HOKUSAI BigWaterfall of the Information Systems Division at RIKEN.
\end{ack}

\bibliographystyle{unsrtnat}
\bibliography{neurips_2021}

\begin{thebibliography}{30}
\providecommand{\natexlab}[1]{#1}
\providecommand{\url}[1]{\texttt{#1}}
\expandafter\ifx\csname urlstyle\endcsname\relax
  \providecommand{\doi}[1]{doi: #1}\else
  \providecommand{\doi}{doi: \begingroup \urlstyle{rm}\Url}\fi

\bibitem[Maldacena(1998)]{Maldacena:1997re}
Juan~Martin Maldacena.
\newblock {The Large N limit of superconformal field theories and
  supergravity}.
\newblock \emph{Adv. Theor. Math. Phys.}, 2:\penalty0 231--252, 1998.
\newblock \doi{10.1023/A:1026654312961}.

\bibitem[Itzhaki et~al.(1998)Itzhaki, Maldacena, Sonnenschein, and
  Yankielowicz]{Itzhaki:1998dd}
Nissan Itzhaki, Juan~Martin Maldacena, Jacob Sonnenschein, and Shimon
  Yankielowicz.
\newblock {Supergravity and the large N limit of theories with sixteen
  supercharges}.
\newblock \emph{Phys. Rev. D}, 58:\penalty0 046004, 1998.
\newblock \doi{10.1103/PhysRevD.58.046004}.

\bibitem[Banks et~al.(1997)Banks, Fischler, Shenker, and
  Susskind]{Banks:1996vh}
Tom Banks, W.~Fischler, S.H. Shenker, and Leonard Susskind.
\newblock {M theory as a matrix model: A Conjecture}.
\newblock \emph{Phys. Rev. D}, 55:\penalty0 5112--5128, 1997.
\newblock \doi{10.1103/PhysRevD.55.5112}.

\bibitem[de~Wit et~al.(1988)de~Wit, Hoppe, and Nicolai]{deWit:1988wri}
B.~de~Wit, J.~Hoppe, and H.~Nicolai.
\newblock {On the Quantum Mechanics of Supermembranes}.
\newblock \emph{Nucl. Phys. B}, 305:\penalty0 545, 1988.
\newblock \doi{10.1016/0550-3213(88)90116-2}.

\bibitem[Berenstein et~al.(2002)Berenstein, Maldacena, and
  Nastase]{Berenstein:2002jq}
David~Eliecer Berenstein, Juan~Martin Maldacena, and Horatiu~Stefan Nastase.
\newblock {Strings in flat space and pp waves from N=4 superYang-Mills}.
\newblock \emph{JHEP}, 04:\penalty0 013, 2002.
\newblock \doi{10.1088/1126-6708/2002/04/013}.

\bibitem[Hanada(2021)]{Hanada:2021ipb}
Masanori Hanada.
\newblock {Bulk geometry in gauge/gravity duality and color degrees of
  freedom}.
\newblock \emph{Phys. Rev. D}, 103\penalty0 (10):\penalty0 106007, 2021.
\newblock \doi{10.1103/PhysRevD.103.106007}.

\bibitem[Berkowitz et~al.(2016)Berkowitz, Rinaldi, Hanada, Ishiki, Shimasaki,
  and Vranas]{Berkowitz:2016jlq}
Evan Berkowitz, Enrico Rinaldi, Masanori Hanada, Goro Ishiki, Shinji Shimasaki,
  and Pavlos Vranas.
\newblock {Precision lattice test of the gauge/gravity duality at large-$N$}.
\newblock \emph{Phys. Rev. D}, 94\penalty0 (9):\penalty0 094501, 2016.
\newblock \doi{10.1103/PhysRevD.94.094501}.

\bibitem[Alford et~al.(1999)Alford, Kapustin, and Wilczek]{Alford:1998sd}
Mark~G. Alford, Anton Kapustin, and Frank Wilczek.
\newblock {Imaginary chemical potential and finite fermion density on the
  lattice}.
\newblock \emph{Phys. Rev. D}, 59:\penalty0 054502, 1999.
\newblock \doi{10.1103/PhysRevD.59.054502}.

\bibitem[Rinaldi et~al.(2021)Rinaldi, Han, Hassan, Feng, Nori, McGuigan, and
  Hanada]{rinaldi2021matrix}
Enrico Rinaldi, Xizhi Han, Mohammad Hassan, Yuan Feng, Franco Nori, Michael
  McGuigan, and Masanori Hanada.
\newblock Matrix model simulations using quantum computing, deep learning, and
  lattice monte carlo, 2021.

\bibitem[Glasser et~al.(2018)Glasser, Pancotti, August, Rodriguez, and
  Cirac]{glasser2018neural}
Ivan Glasser, Nicola Pancotti, Moritz August, Ivan~D Rodriguez, and J~Ignacio
  Cirac.
\newblock Neural-network quantum states, string-bond states, and chiral
  topological states.
\newblock \emph{Physical Review X}, 8\penalty0 (1):\penalty0 011006, 2018.

\bibitem[Carleo et~al.(2019)Carleo, Cirac, Cranmer, Daudet, Schuld, Tishby,
  Vogt-Maranto, and Zdeborov{\'a}]{carleo2019machine}
Giuseppe Carleo, Ignacio Cirac, Kyle Cranmer, Laurent Daudet, Maria Schuld,
  Naftali Tishby, Leslie Vogt-Maranto, and Lenka Zdeborov{\'a}.
\newblock Machine learning and the physical sciences.
\newblock \emph{Reviews of Modern Physics}, 91\penalty0 (4):\penalty0 045002,
  2019.

\bibitem[Saito and Kato(2018)]{saito2018machine}
Hiroki Saito and Masaya Kato.
\newblock Machine learning technique to find quantum many-body ground states of
  bosons on a lattice.
\newblock \emph{Journal of the Physical Society of Japan}, 87\penalty0
  (1):\penalty0 014001, 2018.

\bibitem[Li et~al.(2019)Li, Dee, Khatami, and Johnston]{li2019accelerating}
Shaozhi Li, Philip~M Dee, Ehsan Khatami, and Steven Johnston.
\newblock Accelerating lattice quantum monte carlo simulations using artificial
  neural networks: Application to the holstein model.
\newblock \emph{Physical Review B}, 100\penalty0 (2):\penalty0 020302, 2019.

\bibitem[Nomura(2020)]{nomura2020machine}
Yusuke Nomura.
\newblock Machine learning quantum states—extensions to fermion--boson
  coupled systems and excited-state calculations.
\newblock \emph{Journal of the Physical Society of Japan}, 89\penalty0
  (5):\penalty0 054706, 2020.

\bibitem[Stokes et~al.(2021)Stokes, De, Veerapaneni, and
  Carleo]{stokes2021continuous}
James Stokes, Saibal De, Shravan Veerapaneni, and Giuseppe Carleo.
\newblock Continuous-variable neural-network quantum states and the quantum
  rotor model.
\newblock \emph{arXiv preprint arXiv:2107.07105}, 2021.

\bibitem[Sharir et~al.(2020)Sharir, Levine, Wies, Carleo, and
  Shashua]{sharir2020deep}
Or~Sharir, Yoav Levine, Noam Wies, Giuseppe Carleo, and Amnon Shashua.
\newblock Deep autoregressive models for the efficient variational simulation
  of many-body quantum systems.
\newblock \emph{Physical review letters}, 124\penalty0 (2):\penalty0 020503,
  2020.

\bibitem[Vielhaben and Strodthoff(2021)]{vielhaben2021generative}
Johanna Vielhaben and Nils Strodthoff.
\newblock Generative neural samplers for the quantum heisenberg chain.
\newblock \emph{Physical Review E}, 103\penalty0 (6):\penalty0 063304, 2021.

\bibitem[McNaughton et~al.(2020)McNaughton, Milo{\v{s}}evi{\'c}, Perali, and
  Pilati]{mcnaughton2020boosting}
B~McNaughton, MV~Milo{\v{s}}evi{\'c}, A~Perali, and S~Pilati.
\newblock Boosting monte carlo simulations of spin glasses using autoregressive
  neural networks.
\newblock \emph{Physical Review E}, 101\penalty0 (5):\penalty0 053312, 2020.

\bibitem[Wu et~al.(2019)Wu, Wang, and Zhang]{wu2019solving}
Dian Wu, Lei Wang, and Pan Zhang.
\newblock Solving statistical mechanics using variational autoregressive
  networks.
\newblock \emph{Physical review letters}, 122\penalty0 (8):\penalty0 080602,
  2019.

\bibitem[Albergo et~al.(2019)Albergo, Kanwar, and Shanahan]{albergo2019flow}
MS~Albergo, G~Kanwar, and PE~Shanahan.
\newblock Flow-based generative models for markov chain monte carlo in lattice
  field theory.
\newblock \emph{Physical Review D}, 100\penalty0 (3):\penalty0 034515, 2019.

\bibitem[Medvidovic et~al.(2020)Medvidovic, Carrasquilla, Hayward, and
  Kulchytskyy]{medvidovic2020generative}
Matija Medvidovic, Juan Carrasquilla, Lauren~E Hayward, and Bohdan Kulchytskyy.
\newblock Generative models for sampling of lattice field theories.
\newblock \emph{arXiv preprint arXiv:2012.01442}, 2020.

\bibitem[Nicoli et~al.(2021)Nicoli, Anders, Funcke, Hartung, Jansen, Kessel,
  Nakajima, and Stornati]{nicoli2021estimation}
Kim~A Nicoli, Christopher~J Anders, Lena Funcke, Tobias Hartung, Karl Jansen,
  Pan Kessel, Shinichi Nakajima, and Paolo Stornati.
\newblock Estimation of thermodynamic observables in lattice field theories
  with deep generative models.
\newblock \emph{Physical review letters}, 126\penalty0 (3):\penalty0 032001,
  2021.

\bibitem[Kanwar et~al.(2020)Kanwar, Albergo, Boyda, Cranmer, Hackett,
  Racaniere, Rezende, and Shanahan]{kanwar2020equivariant}
Gurtej Kanwar, Michael~S Albergo, Denis Boyda, Kyle Cranmer, Daniel~C Hackett,
  S{\'e}bastien Racaniere, Danilo~Jimenez Rezende, and Phiala~E Shanahan.
\newblock Equivariant flow-based sampling for lattice gauge theory.
\newblock \emph{Physical Review Letters}, 125\penalty0 (12):\penalty0 121601,
  2020.

\bibitem[Cao et~al.(2019)Cao, Titov, and Aziz]{decao2019block}
Nicola~De Cao, Ivan Titov, and Wilker Aziz.
\newblock Block neural autoregressive flow, 2019.

\bibitem[Han and Hartnoll(2020)]{Han:2019wue}
Xizhi Han and Sean~A. Hartnoll.
\newblock {Deep Quantum Geometry of Matrices}.
\newblock \emph{Phys. Rev. X}, 10\penalty0 (1):\penalty0 011069, 2020.
\newblock \doi{10.1103/PhysRevX.10.011069}.

\bibitem[Luo et~al.(2020)Luo, Carleo, Clark, and Stokes]{luo2020gauge}
Di~Luo, Giuseppe Carleo, Bryan~K Clark, and James Stokes.
\newblock Gauge equivariant neural networks for quantum lattice gauge theories.
\newblock \emph{arXiv preprint arXiv:2012.05232}, 2020.

\bibitem[Luo et~al.(2021)Luo, Chen, Hu, Zhao, Hur, and Clark]{luo2021gauge}
Di~Luo, Zhuo Chen, Kaiwen Hu, Zhizhen Zhao, Vera~Mikyoung Hur, and Bryan~K
  Clark.
\newblock Gauge invariant autoregressive neural networks for quantum lattice
  models.
\newblock \emph{arXiv preprint arXiv:2101.07243}, 2021.

\bibitem[Favoni et~al.(2020)Favoni, Ipp, M{\"u}ller, and
  Schuh]{favoni2020lattice}
Matteo Favoni, Andreas Ipp, David~I M{\"u}ller, and Daniel Schuh.
\newblock Lattice gauge equivariant convolutional neural networks.
\newblock \emph{arXiv preprint arXiv:2012.12901}, 2020.

\bibitem[K{\"o}hler et~al.(2020)K{\"o}hler, Klein, and
  No{\'e}]{kohler2020equivariant}
Jonas K{\"o}hler, Leon Klein, and Frank No{\'e}.
\newblock Equivariant flows: exact likelihood generative learning for symmetric
  densities.
\newblock In \emph{International Conference on Machine Learning}, pages
  5361--5370. PMLR, 2020.

\bibitem[Witten and Olive(1978)]{Witten:1978mh}
Edward Witten and David~I. Olive.
\newblock {Supersymmetry Algebras That Include Topological Charges}.
\newblock \emph{Phys. Lett. B}, 78:\penalty0 97--101, 1978.
\newblock \doi{10.1016/0370-2693(78)90357-X}.

\end{thebibliography}

\end{document}